\documentclass{jetpl}
\usepackage{color,amsmath,amsfonts,graphics,epsfig,amssymb}
\usepackage[english]{babel}
\usepackage{hyperref}
\twocolumn
\lat

\title{
\begin{flushright}
\normalsize \rm INR-TH-2022-021
\end{flushright}
\vspace{8mm}
Parameters of axion-like particles required to explain high-energy photons from GRB~221009A
}

\rtitle{Axion-like particles and GRB~221009A}

\sodtitle{Parameters of axion-like particles required to explain high-energy photons from GRB~221009A}

\author{S.\,V.\,Troitsky\thanks{e-mail: st@ms2.inr.ac.ru}}

\rauthor{S.\,V.\,Troitsky}

\sodauthor{S.\,V.\,Troitsky}

\address{Institute for Nuclear Research of the Russian Academy of
Sciences,\\
60th October Anniversary prospect 7A, 117312 Moscow, Russia}

\dates{October 17, 2022}{*}

\abstract{
Recent astrophysical transient Swift~J1913.1+1946 is possibly associated with the gamma-ray burst GRB~221009A at the redshift $z\approx 0.151$. The transient was accompanied by very high-energy gamma rays up to 18~TeV observed by LHAASO and a photon-like air shower of 251~TeV detected by Carpet-2. These energetic gamma rays cannot reach us from the claimed distance of the source because of the pair production on cosmic background radiation. If the identification and redshift measurements are correct, one would require new physics to explain the data. One possibility invokes axion-like particles (ALPs) which mix with photons but do not attenuate on the background radiation. Here we explore the ALP parameter space and find that the ALP--photon mixing in the Milky Way, and not in the intergalactic space, may help to explain the observations. However, given the low Galactic latitude of the event, misidentification with a Galactic transient remains an undiscarded explanation.}
\begin{document}
\maketitle

\noindent\textbf{1. Observations.} An unusual energetic astrophysical transient Swift~J1913.1+1946 has been detected on October 9, 2022 \cite{Swift-GCN} and soon associated with a gamma-ray burst GRB~221009A detected by Fermi GBM \cite{FermiGBM-GCN}. The redshift of the GRB was determined from absorption lines in the afterglow \cite{z-VLT-GCN,z-GTC-GCN} and from emission lines in the host galaxy \cite{z-host-VLT-GCN} as $z\approx 0.151$. The main peculiarity of the transient is the presence of extremely energetic gamma rays, never detected from a GRB. In particular, LHAASO reported the detection of thousands of photons with energies up to 18~TeV in the first 2000~s after the GRB trigger \cite{LHAASO-GCN}, and Carpet-2 reported the detection of a 251-TeV photon-like air shower 4536~s after the trigger \cite{CarpetATel-GRB}. We assume, for the main part of the present Letter, that the photons observed by LHAASO and Carpet-2 indeed arrived from the GRB at the reported distance. Potential misidentifications will be briefly addressed in Sec.~4.

\vskip 1mm
\noindent\textbf{2. Photon attenuation and axion-like particles.}
The observations by both LHAASO and Carpet-2 challenge conventional understanding because gamma rays of that high energies cannot reach us from distant sources \cite{Nikishov}. They should instead produce $e^+e^-$ pairs on the cosmic background radiation. The predicted mean free path for 18-TeV photons depends on the assumed extragalactic infrared background which is known with considerable uncertainties, so the optical depth for a source at $z=0.151$ is estimated as $\sim 15 \pm 5$. A photon with the energy of 251~TeV produce pairs on much more abundant, and better known, cosmic microwave background (CMB), and its mean free path is only of order 75~kpc, smaller than the virial radius of our own Milky Way Galaxy; the optical depth for the distance of GRB~221009A is $>3000$ for this energy.

The problem of observations of energetic particles, possibly photons, from very distant sources arose in various contexts and always required non-standard physics to be solved or relaxed. Here we concentrate on mixing of photons with hypothetical axion-like particles (ALPs) in the external magnetic field \cite{RaffeltStodolsky}, which was first invoked to solve an astrophysical problem of this kind in Ref.~\cite{Csaba}, and for the gamma-ray propagation in Ref.~\cite{Roncadelli-2007}. For reviews and more references, see e.g.\ \cite{ST-mini-rev,Tinyakov:2021lnt,Roncadelli-review2022}.

A general ALP is characterized by two parameters, mass $m$ and photon coupling $g$. The latter appears in the specific interaction term in the Lagrangian, $-(g/4)aF_{\mu\nu}\tilde F^{\mu\nu}$, where $a$ is the pseudoscalar ALP field, $F_{\mu\nu}$ is the electromagnetic stress tensor and $\tilde F^{\mu\nu}=(1/2)\epsilon^{\mu\nu\rho\lambda}F_{\rho\lambda}$ is its dual tensor. Compared to the QCD axion, ALP is more general because $m$ and $g$ are arbitrary independent parameters, but at the same time is simpler because it does not necessarily interact with gluons in the minimal version. The ALP/photon mixing is described in Ref.~\cite{RaffeltStodolsky}, while a collection of equations relevant in the astrophysical context may be found e.g.\ in Ref.~\cite{FRT:2009}, but also in many other papers and textbooks; we do not repeat them here.

ALPs do not produce pairs and so they propagate unattenuated through the Universe, so the ALP/photon mixing makes the following scenario possible: passing through sufficiently strong magnetic fields, photons convert to ALPs and back, and hence the mixed particle beam can travel longer than pure photons. For extragalactic astrophysics, it is important to distinguish two particular cases, see discussions in Ref.~\cite{ST-2regimes}. In the first limit, ALP parameters allow for conversion in the (rather weak) extragalactic magnetic field, so the photon-ALP oscillations happen along the entire path from the source to the observer \cite{Csaba,Roncadelli-2007}. In the opposite case, stronger fields are required for non-negligible mixing, so that the conversion happens near the source, in the host galaxy, cluster or filament, and again in the Local Supercluster or in the Milky Way \cite{Serpico,FRT:2009}. For very distant sources, this difference is crucial. In the case of intergalactic mixing, the gamma-ray part of the mixed beam is constantly fed by the ALP part and attenuates, so, in the limit of large distance, all energy finally goes to $e^+e^-$ pairs. This means that the mean free path becomes effectively longer by a certain factor. If the intergalactic mixing is suppressed, then some part of photons may convert to ALPs near the source and reconvert back to gamma rays near the observer; the remaining gamma-ray part of the beam attenuates as usual. In the limit of large distances, larger photon fluxes are expected to be observed in the latter case; see Ref.~\cite{ST-2regimes} for details. Here we concentrate on this latter case only.

\vskip 1mm
\noindent\textbf{3. ALPs and GRB~221009A.} 
In the last few days, immediately after the observations of high-energy photons possibly associated with GRB~221009A had been reported, explanation of these observations in terms of ALP/photon mixing was addressed in Refs.\ 
\cite{Roncadelli-newGRB} and \cite{Meyer-newGRB}. In Ref.~\cite{Roncadelli-newGRB}, the scenario of intergalactic mixing was discussed, assuming rather strong intergalactic magnetic field of 1~nG. The resulting photon survival probability was low, $\sim 10^{-5}$, in agreement with qualitative reasons explained above. In Ref.~\cite{Meyer-newGRB}, the opposite scenario was discussed, but for the LHAASO 18-TeV photons only. Here we present a joint analysis of both observations and determine the favored range of ALP parameters.

At the time of this writing, little is known about the source of GRB~221009A and its host galaxy, so we simply assume the maximal mixing, that is equipartition between two photon polarizations and ALP, in the source. This assumption is backed up by the fact that the magnetic fields in the host galaxy and in our Galaxy are expected to be similar, while we are looking for the mixing close to maximal in the Milky Way. Mixing smaller than maximal would result in shifting the relevant parameter space to higher values of $g$. Next, we neglect the mixing in the intergalactic space to avoid the corresponding flux suppression discussed above. As a result, we assume that the state arriving to the Milky Way is pure ALP and the flux is 1/3 of the emitted photon flux. For the propagation of the ALP-photon beam in the Milky Way, we solve numerically the evolution equation in the density-matrix formalism, as described e.g.\ in Ref.~\cite{LT-magnetic-fields}. 
We use the Galactic magnetic field model of Ref.~\cite{Pshirkov-GMF} for the line of sight corresponding to celestial coordinates of GRB~221009A. The initial condition for the density matrix is $\rho(y_0)=\mbox{diag}(0,0,1/3)$ at the coordinate $y_0$ corresponding to the distance of 20~kpc from the Galactic Center (the magnetic field of the model we use is zero beyond this point). We note that the direction to GRB~221009A has supergalactic latitude $b_{\rm SG}\approx 84^\circ$, hence the magnetic field of the Local Supercluster may be safely neglected. The survival probability of the photon is given by $\rho_{11}(0)+\rho_{22}(0)$, where $y=0$ corresponds to the location of the observer.

\begin{figure}
    \centering
    \includegraphics[width=0.9\linewidth]{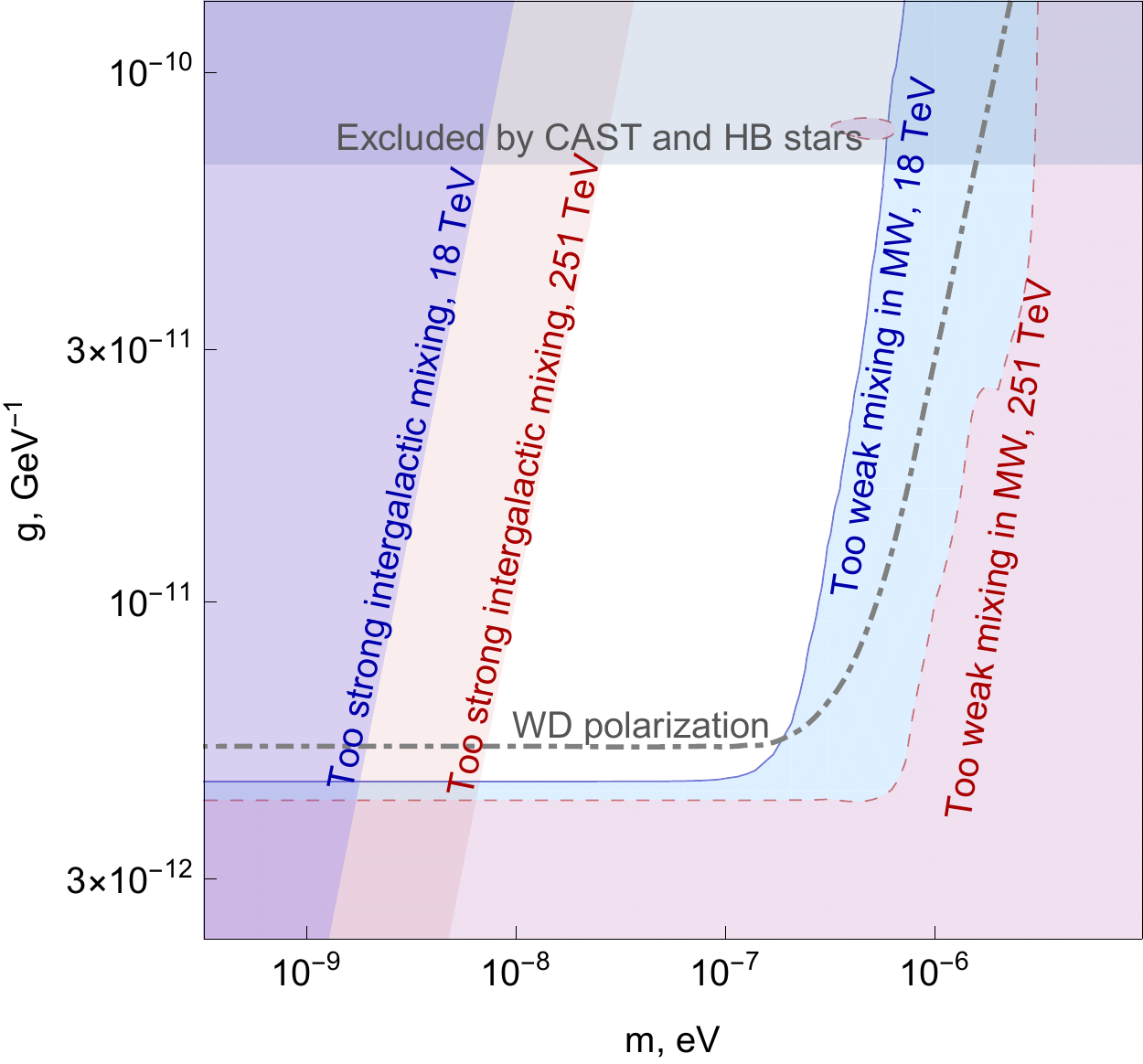}
    \caption{\label{fig:main} \sl
\textbf{Figure~\ref{fig:main}.}
ALP mass $m$ and photon coupling $g$. ALPs with parameters in the white central part of the plot can explain both 18-TeV and 251-TeV photons ($\ge 1\%$ of photons survive assuming maximal mixing in the source and reconversion in the Milky Way, MW, for the direction of GRB~221009A, the Galactic magnetic field model of Ref.~\cite{Pshirkov-GMF}). The top band is disfavoured by the CAST search for solar axions \cite{CAST} and by constraints from the evolution of horizontal-branch (HB) stars \cite{HBstars}. In the shaded area to the left, strong mixing in the intergalactic space (assuming the magnetic field of 1~nG) results in strong suppression of the photon flux. The upper limit from magnetic white dwarf polarization \cite{MWD-polarization} is shown by the gray dash-dotted line.
See the text for details and discussion.}
\end{figure}
Our main results are collected in Figure~\ref{fig:main}, which presents the ALP parameter space. Above the full blue and dashed red lines, the surviving probability for photons of 18~TeV and 251~TeV, respectively, exceeds 1\%, so that it makes sense to hope that the ALP-gamma conversion could help to observe gamma rays from the distant source. Note that a more quantitative study should use estimated gamma-ray fluxes which have not been published by LHAASO, nor by Carpet-2. The acceptance of these surface arrays depends strongly on selection cuts used in the analysis and on the zenith angle under which the source is observed, so it might be misleading to infer the exposure from other publications or to use long-term exposure averaged over zenith angles to estimate the flare flux.

To justify our approximation and to avoid strong flux suppression, we need to guarantee that the mixing in intergalactic magnetic field $B_{\rm IG}$ is suppressed. This is reached for \cite{FRT:2009}
$$
\frac{g}{10^{-11}~\mbox{GeV}^{-1}}
\lesssim
\left(\frac{E}{255~\mbox{TeV}}\right)^{-1}
\left(\frac{m}{10^{-8}~\mbox{eV}}\right)^{2}
\left(\frac{B_{\rm IG}}{\rm nG}\right)^{-1}.
$$
The regions where this condition is not satisfied are shown in Fig.~\ref{fig:main} for two values of $E$ as shaded regions to the left. Note that the value of 1~nG used for presenting this estimate in the plot is close to the observational upper bound on the intergalactic magnetic field \cite{Pshirkov-IGMF}, while its actual value may be orders of magnitude smaller; the lines would shift to smaller $m$ in this case. The remaining white central part of the plot corresponds to the values of $m$ and $g$ for which the observations by LHAASO and Carpet-2 may be explained by photon-ALP mixing. These parameters are in fact motivated in some ALP models, e.g.\ \cite{2210.08841,2210.10022}.

\vskip 1mm
\noindent\textbf{4. Discussion.}

\vskip 0.5mm
\noindent{\sl 4.1. Other constraints on ALPs.}
Among various experimental constraints on the ALP parameters, the most relevant here are the upper bounds on $g$ from non-observation of solar axions by the CAST experiment \cite{CAST} and from the study of evolution of helium-burning stars in globular clusters \cite{HBstars}, coincident numerically by chance. The shaded area in the upper part of Fig.~\ref{fig:main} represents these constraints. In addition, there exist constraints from various astrophysical photon observations which probe the part of the ALP parameter space we discuss here; however, they depend on the assumptions about poorly known magnetic fields in astrophysical sources. This dependence may introduce huge systematic uncertainties in the resulting constraints, see e.g.\ Ref.~\cite{LT-magnetic-fields,Tinyakov:2021lnt}. The strongest of these constraints comes from the polarization of a magnetic white dwarf SDSS~J135141 \cite{MWD-polarization} and results in a $m$-dependent upper limit on $g$ shown in Fig.~\ref{fig:main} as a gray dash-dotted line (fiducial upper bound from Fig.~1 of Ref.~\cite{MWD-polarization}). Interestingly, even with the account of this strict bound, there remains an allowed part of the parameter space for the explanation of the observed energetic photons.

\vskip 0.5mm
\noindent{\sl 4.2. Probability of the background events.}
Most of the events detected by surface air-shower arrays, like LHAASO and Carpet-2, are caused by charged cosmic rays, and only a small fraction of them are photons. Even with strict selection cuts imposed, there exists a nonzero probability of misidentification of a cosmic-ray shower as a photon-induced one. For LHAASO, Ref.~\cite{Meyer-newGRB} estimated the expected number of background cosmic-ray events with $\sim 18$~TeV energies during 2000~sec observation time as 2.8, using published results of a different LHAASO analysis as the input. We note that, like the exposure, this background rate depends strongly on the used selection cuts, which may vary from one analysis to another, and on the zenith angle, so the real value may differ significantly from this estimate. It is hard to estimate the probability of chance coincidence of background events with the GRB without detailed knowledge of the selection procedure and of the energy distribution of observed events.

On the other hand, Carpet-2 events similar to the 251-TeV photon-like shower are certainly rare. Carpet-2 is a relatively small installation, and even a huge flare of a Galactic source resulted in the detection of only several tens of events there \cite{Carpet-ApJL2021}. For the observed event associated with GRB~221009A, the probability of the background coincidence of $1.2 \times 10^{-4}$ was reported by the experiment \cite{CarpetATel-GRB}. 

\vskip 0.5mm
\noindent{\sl 4.3. Galactic sources.}
Even for the reported temporal and directional coincidence of detected photons with GRB~221009A, there remains a possibility that the highest-energy events came from a Galactic source, especially given the low Galactic latitude, $b\approx 4^\circ$, of the event \cite{Carpet-ApJL2021}. This would solve the attenuation problem automatically, because the photons would not have time to produce pairs even on CMB. No particularly interesting, or known as flaring, Galactic source coincides with the GRB location, known with the arcsecond precision. For the Carpet-2 angular resolution of several degrees, two previously reported, possibly coinciding Galactic sources of photons above 100~TeV, namely 3HWC~J1928$+$178 and LHAASO~J1929$+$1745, are located not far from the direction to GRB~221009A in the sky \cite{HAWCATel-Carpet-sources}. It remains to be understood whether their activity can be responsible for the observed events.

It can even be possible that a superposition of a GRB and a Galactic flare was observed. To some extent this may even explain the unusual light curve of the transient \cite{FermiLAT-GCN,Konus-GCN}. All these possibilities remain to be explored in future work, when more data will be accumulated and analysed jointly.

\vskip 1mm
\noindent\textbf{5. Summary.}
To summarize, the ALP explanation of exotic very energetic photons from GRB~221009A is viable: the probability to detect such photons is considerable for ALP mass and photon coupling in an allowed region of the parameter space, see Fig.~\ref{fig:main}. The conversion should happen in the host galaxy and in the Milky Way. Because of a very short mean free path of $\sim 250$~TeV gamma rays, the observation of a photon of this energy from a distant source disfavours the explanation involving ALP/$\gamma$ mixing in intergalactic space. Future careful work should be performed, however, to test the possibility of misidentification of the GRB, or of some of photons associated with it, with a flare of a Galactic source.

\vskip 1mm
\noindent\textbf{Acknowledgements.} The author is indebted to T.~Dzhatdoev,  E.~Podlesny, V.~Rubakov and G.~Rubtsov for interesting and helpful discussions and important comments on the manuscript. This work was supported by the Russian Science Foundation, grant 22-12-00253.

\vskip 1mm
\noindent\textbf{Dedication.} I dedicate this work to the memory of my teacher Valery Rubakov, a great physicist and a great personality. Few hours before submission of this Letter, he gave me his comments on the draft, and a day later he passed away unexpectedly. We will miss him forever.
\bibliographystyle{nature}
\bibliography{grb}
\end{document}